\documentclass[pra,english,twocolumn,aps]{revtex4-1}
\usepackage[T1]{fontenc}
\usepackage[latin9]{inputenc}
\usepackage{xcolor}
\usepackage{amsmath}
\usepackage{amssymb}
\usepackage{graphicx}
\usepackage{natbib}
\usepackage{epstopdf}
\usepackage[colorlinks=true,linkcolor=blue,citecolor=blue,urlcolor=blue]{hyperref}

\makeatletter

\@ifundefined{textcolor}{}
{%
 \definecolor{BLACK}{gray}{0}
 \definecolor{WHITE}{gray}{1}
 \definecolor{RED}{rgb}{1,0,0}
 \definecolor{GREEN}{rgb}{0,1,0}
 \definecolor{BLUE}{rgb}{0,0,1}
 \definecolor{CYAN}{cmyk}{1,0,0,0}
 \definecolor{MAGENTA}{cmyk}{0,1,0,0}
 \definecolor{YELLOW}{cmyk}{0,0,1,0}
}

\makeatother

\usepackage{babel}
\begin{document}
\title{Criticality-Based Quantum Metrology in the Presence of Decoherence}

\author{Wan-Ting He}
\affiliation{Department of Physics, Applied Optics Beijing Area Major Laboratory,
Beijing Normal University, Beijing 100875, China}

\author{Cong-Wei Lu}
\affiliation{Department of Physics, Applied Optics Beijing Area Major Laboratory, Beijing Normal University, Beijing 100875, China}

\author{Yi-Xuan Yao}
\affiliation{Department of Physics, Applied Optics Beijing Area Major Laboratory, Beijing Normal University, Beijing 100875, China}

\author{Hai-Yuan Zhu}
\affiliation{Department of Physics, Applied Optics Beijing Area Major Laboratory, Beijing Normal University, Beijing 100875, China}

\author{Qing Ai}
\email{aiqing@bnu.edu.cn}
\affiliation{Department of Physics, Applied Optics Beijing Area Major Laboratory, Beijing Normal University, Beijing 100875, China}

\date{\today}

\begin{abstract}
Quantum metrology aims to use quantum resources to improve the precision
of measurement. Because quantum systems with criticality are very
sensitive to the variation of order parameters around the quantum
phase transition (QPT), quantum criticality has been presented as
a novel and efficient resource. Generally, protocols
of criticality-based quantum metrology often work without decoherence.
In this paper, we address the issue whether the divergent feature
of the inverted variance is indeed realizable in the presence of noise
when approaching the QPT. Taking the quantum Rabi model (QRM) as an example, we obtain the analytical result for the inverted variance. 
We show that the inverted variance may
be convergent in time due to the noise. When approaching the critical point, the maximum inverted variance demonstrates a power-law increase with the exponent $-1.2$, of which the absolute value is smaller than that for the noise-free case, i.e., $2$. We also observe a power-law dependence of the maximum inverted variance on the relaxation rate and the temperature. Since the precision of the metrology is very sensitive to the noise, as a remedy, we propose performing the squeezing operation on the initial state to improve the precision under decoherence. In addition, we also investigate the criticality-based metrology under the influence of the two-photon relaxation. Contrary to the single-photon relaxation, the quantum dynamics of the inverted variance shows a completely-different behavior. It does not oscillate with the same frequency with respect to the re-scaled time for different dimensionless coupling strengths. Strikingly, although the maximum inverted variance still manifests a power-law dependence on the energy gap, the exponent is positive and depends on the dimensionless coupling strength. This observation implies that the criticality may not enhance but weaken the precision in the presence of two-photon relaxation. It can be well described by the non-linearity introduced by the two-photon relaxation. 
\end{abstract}

\maketitle

\section{Introduction}

Quantum criticality has been shown to provide significant advantages
for quantum sensing and metrology \citep{Heugel2019PRL,Chu2021PRL,Rams2018PRX,Garbe2020PRL,Felicetti2020PRL,Ilias2022PRXQ,Zanardi2008PRA,Tsang2013PRA,Fern2017PRA,Gietka2021Quantum}.
In quantum phase transition (QPT), a small variation in the order
parameter may cause a huge change in the system's properties when
approaching the quantum critical point \citep{Sachdev2011,Quan2006PRL,Ai2009SCSG}. Thus,
quantum criticality can be viewed as a valuable resource to acquire
an ultrahigh-precision estimation. In the last few years, criticality-based
quantum metrological schemes have gained increasing attention among
both theoretical and experimental researchers. Many criticality-based
quantum metrology protocols in QPT systems have been proposed, which
can be classified into two types \citep{Felicetti2020PRL,Garbe2020PRL}.
One of them focuses on the critical behaviors of the QPT at equilibrium,
and utilizes the ground state of the Hamiltonian near the critical
point. Unfortunately, it is very inefficient to prepare the ground-state
around the critical point by the state-of-art technology. The other
focuses on the dynamical behaviors, which evolve under a Hamiltonian
close to the critical point. It is quite similar to the interferometric paradigm in the typical quantum metrology. Previous work \citep{Rams2018PRX} has revealed that both two kinds of approaches yield the same scaling
for a broad class of quantum many-body systems. Recently,
a dynamic framework for criticality-based quantum metrology, which
preliminarily focuses on the dynamical approach, has been proposed
\citep{Chu2021PRL,Pang2017NC,Pang2014PRA}. The fundamental idea
as well as the details of experimental implementation using quantum
Rabi model (QRM) have been illustrated. The QRM is one of the most
fundamental models describing quantum light-matter interactions \cite{Scully1997}. There
exists a superradiant QPT in the thermodynamic limit of the QRM \citep{Hwang2015PRL,Puebla2017PRL,Lv2018PRX,Pedernales2015SR}.
The thermodynamic limit can be achieved when the reduced Rabi frequency
approaches infinity in some quantum systems with criticality, such
as the Dicke model, the Lipkin-Meshkov-Glick model and the QRM \citep{Felicetti2020PRL}.

On the other hand, any quantum system inevitably suffers from the
interaction with the surrounding environment and thus forms an open
quantum system. It has been known over two decades that the quantum
metrology with the maximum-entangled states shows no advantage over the classical counterpart when the
decoherence is present \citep{Huelga1997PRL}. Interestingly, when
placed in a non-Markovian environment, the entangled probes indeed
manifest their superiority by the quantum Zeno effect \citep{Chin2012PRL,Kofman2012PR,Ai2010PRA,Ai2013SR,Harrington2017PRL},
and recently it was experimentally demonstrated by an exact and efficient quantum simulation approach \citep{Long2022PRL,Buluta2009S,Georgescu2014RMP,Zhang2020FoP,Wang2018NPJQI,Chen2022NPJQI}.

In this work, inspired by the above discoveries,
we analyze the achievable precision in the QRM when considering the
thermal relaxation. The upper bound on the achievable precision of
the metrology will be dramatically reduced owing to the noise. This result
indicates that the dissipation poses severe limitations and thus hinders
the metrological advantages in this framework. Furthermore, we find
that the inverted variance may converge in time no matter as a result of
the noise. 
We further explore optimization of initial
state and find that the precision can be improved by a squeezing operation
on the initial state. On the other hand, recently two-photon relaxation has been experimentally realized \cite{Leghtas2015Science}. It was theoretically shown that the two-photon relaxation preserves the $Z_2$ symmetry of the QRM. The initial states with even-odd parity manifest qualitatively
distinct transient and steady state behaviors, which is present both at ultrastrong and weak couplings \cite{Malekakhlagh2019PRL}. Motivated by these observations, we theoretically investigate the quantum dynamics of the metrology in the presence of two-photon relaxation.

This paper is organized as follows. In Sec.~\ref{sec:Method}, we
outline some basic concepts in the quantum metrology based on the criticality
and introduce the QRM with noise. In Sec.~\ref{sec:SinglePhoton}, we analyze
the results by the semiclassical equation of motion and discuss the effects of various physical parameters on the measurement precision. In Sec.~\ref{sec:TwoPhoton}, we investigate the effects of the two-photon relaxation on the metrology. The main conclusions of this paper are drawn in Sec.~\ref{sec:Conclusion}.

\section{Model}\label{sec:Method}

In this paper, we consider the QRM for criticality-based quantum metrology,
which consists of a two-level system coupled to a single cavity mode.
It is the most simplified version of the Dicke model \citep{Dicke1954PR}.
The Hamiltonian of the QRM is
\begin{equation}
H_{\text{Rabi}}=\omega a^{\dagger}a+\frac{\omega_{0}}{2}\sigma_{z}-\lambda(a+a^{\dagger})\sigma_{x},
\end{equation}
where $\omega$ is the frequency of the bosonic field, $a^{\dagger}$
and $a$ are the creation and annihilation operators of the field,
$\omega_{0}$ is the transition frequency of the two-level atom, $\sigma_{\alpha}$
$(\alpha=x,y,z)$ are the Pauli operators of the atom, $\lambda$
is the coupling strength. Let $g=2\lambda/\sqrt{\omega\omega_{0}}$
be the dimensionless coupling constant. When the ratio of two transition
frequencies diverges, i.e., $\eta=\omega_{0}/\omega\rightarrow\infty$,
the energy gap closes and there exhibits a second-order normal-to-superradiant
QPT at the critical point $g=1$ \citep{Hwang2015PRL}. In the limit
of $\eta\rightarrow\infty$, through the Schrieffer-Wolff transformation
$U_{\text{np}}=\exp[g\sqrt{\eta^{-1}}(a+a^{\dagger})(\sigma_{+}-\sigma_{-})/2]$
\citep{Schrieffer1966PR}, we can obtain an effective low-energy normal-phase
Hamiltonian as
\begin{equation}
H_{\text{np}}=\omega[P^{2}+(1-g^{2})X^{2}]/2,
\end{equation}
where $X=(a+a^{\dagger})/\sqrt{2}$ and $P=i(a^{\dagger}-a)/\sqrt{2}$
are the quadrature operators.
Let $H_{0}=\omega P^{2}/2$ and $H_{1}=\omega X^{2}/2$. The quantum Fisher information (QFI) for the estimation of the parameter $g$ around the critical point can be expressed as \citep{Chu2021PRL}
\begin{equation}
\mathcal{F}_{g}\simeq16g^{2}\frac{\left[\sin(\sqrt{\Delta_{g}}\omega t)-\sqrt{\Delta_{g}}\omega t\right]^{2}}{\Delta_{g}^{3}}\text{Var}(P^{2})_{|\psi\rangle},
\end{equation}
where $\Delta_{g}=4(1-g^{2})$ characterizes the energy gap, $\text{Var}(P^{2})_{|\psi\rangle}=\langle\psi\vert P^{2}|\psi\rangle-\langle\psi\vert P|\psi\rangle^{2}$
is the variance of the momentum in the state $|\psi\rangle$. In experiments,
the precision of measurement is characterized by the inverted variance
as $F_{g}=\chi_{g}^{2}/(\Delta X)^{2}$ with $\chi_{g}=\partial_{g}\langle X\rangle$ and $(\Delta X)^{2}=\langle X^{2}\rangle-\langle X\rangle^{2}$. Here, $\langle X\rangle$ is the expectation of $X$. Assume
an initial state $|\psi(0)\rangle=|\downarrow\rangle\otimes|\psi\rangle_{b}$
with the spin in the spin-down state $|\downarrow\rangle$ and the bosonic field state $|\psi\rangle_{b}=(|0\rangle+i|1\rangle)/\sqrt{2}$.
The inverted variance will achieve its local maxima $F_{g}(\tau)=\chi_{g}^{2}/(\Delta X)^{2}|_{t=\tau}=8g^{2}\omega^{2}\Delta_{g}^{-2}\tau^{2}$
at $\tau=2m\pi/(\sqrt{\Delta_{g}}\omega)$ with $m\in\mathbb{Z}$. Notice that the inverted variance diverges in the long-time limit.
The QFI at the same time is $\mathcal{F}_{g}(\tau)\simeq16g^{2}\omega^{2}\Delta_{g}^{-2}\tau^{2}\text{Var}(P^{2})_{|\psi\rangle_{b}}$.
The local maximum of the inverted variance $F_{g}(\tau)$ is of the
same order of the QFI $\mathcal{F}_{g}(\tau)$.

\section{Single-Photon Relaxation}\label{sec:SinglePhoton}

First of all, we consider the QRM with the single-photon relaxation.
The inverted variance can be obtained solving the Lindblad-form
master equation numerically by using QuTip \cite{Breuer2002,Johansson2012CPC,Johansson2013CPC}. In the $\eta\rightarrow\infty$
limit, all the corrections which have an order higher than $\eta^{-1/2}$
become zero. Thus, we have
\begin{align}
U_{\textrm{np}}^{\dagger}aU_{\textrm{np}} & \approx a.
\end{align}
Upon the projection onto the spin-down subspace, we obtain the effective
master equation $\dot{\rho}=-i[H_{\text{np}},\rho]+\gamma_{a}D[a]+\gamma_{h}D[a^{\dagger}]$,
where $D[a]=a\rho a^{\dagger}-a^{\dagger}a\rho/2-\rho a^{\dagger}a/2$,
$\gamma_{a}$ is the decay rate, $\gamma_{h}$ is the heating rate
\citep{Carmichael1993}. Applying a semiclassical equation of motion
for the open QRM \cite{Gardiner2004,Hwang2018PRA}, we have
\begin{eqnarray}
\frac{d}{dt}\langle X\rangle & = & -\frac{\gamma_{a}-\gamma_{h}}{2}\langle X\rangle+\omega\langle P\rangle,\nonumber \\
\frac{d}{dt}\langle P\rangle & = & -\frac{\gamma_{a}-\gamma_{h}}{2}\langle P\rangle-\frac{\Delta_{g}\omega}{4}\langle X\rangle,\nonumber \\
\frac{d}{dt}\langle X^{2}\rangle & = & -(\gamma_{a}-\gamma_{h})\left\langle X^{2}\right\rangle +\omega\langle G\rangle+\frac{\gamma_{a}+\gamma_{h}}{2},\label{eq:SemEq}\\
\frac{d}{dt}\langle P^{2}\rangle & = & -(\gamma_{a}-\gamma_{h})\left\langle P^{2}\right\rangle -\frac{\Delta_{g}\omega}{4}\langle G\rangle+\frac{\gamma_{a}+\gamma_{h}}{2},\nonumber \\
\frac{d}{dt}\langle G\rangle & = & -(\gamma_{a}-\gamma_{h})\langle G\rangle+2\omega\langle P^{2}\rangle-\frac{\Delta_{g}\omega}{2}\langle X^{2}\rangle,\nonumber
\end{eqnarray}
where $\langle\cdot\rangle=\text{Tr}(\cdot\rho_{b})$ with $\rho_{b}$
being the density matrix of the bosonic field and $G=XP+PX$. In general,
$\gamma_{a}=\kappa(\bar{n}+1)$ and $\gamma_{h}=\kappa\bar{n}$ describe
the coupling of the field to a thermal reservoir at temperature $T$
with mean number
\begin{equation}
\bar{n}=\frac{1}{e^{\omega/k_BT}-1}
\end{equation}
of thermal photons, $k_B$ the Boltzmann constant, $\kappa$ the decay
rate at zero temperature \citep{Carmichael1993}. By solving Eq.~(\ref{eq:SemEq}),
the inverted variance $F_{g}(t)$ is analytically given as
\begin{equation}
F_{g}(t)=\frac{4\Delta_{g}^{-2}(4-\Delta_{g})A^{2}(t)}{B(t)+\frac{2n+1}{\Delta_{g}\omega^{2}+\kappa^{2}}C(t)},\label{eq:Fgt}
\end{equation}
where the three time-dependent factors are respectively
\begin{eqnarray}
A(t) && =[\Delta_{g}\omega t\langle X(0)\rangle+4\langle P(0)\rangle]\sin(\frac{1}{2}\sqrt{\Delta_{g}}\omega t)\nonumber \\
 &&\;\;\;\; -2\sqrt{\Delta_{g}}\omega t\langle P(0)\rangle\cos(\frac{1}{2}\sqrt{\Delta_{g}}\omega t),\\
B(t) && =  2\Delta_{g}[1+\cos(\sqrt{\Delta_{g}}\omega t)]\text{Var}(X^{2})_{|\psi(0)\rangle}\nonumber \\
 &&\;\;\;\; +8[1-\cos(\sqrt{\Delta_{g}}\omega t)]\text{Var}(P^{2})_{|\psi(0)\rangle}\nonumber \\
 &&\;\;\;\; +8\sqrt{\Delta_{g}}\sin(\sqrt{\Delta_{g}}\omega t)\textrm{Re}[\mathrm{Cov}(X,P)],\\
C(t) && =\Delta_{g}(\Delta_{g}\omega^{2}+4\omega^{2}+2\kappa^{2})(e^{\kappa t}-1)-(4-\Delta_{g})\times\nonumber \\
 && \{ \kappa^{2}[1-\cos(\sqrt{\Delta_{g}}\omega t)]+\sqrt{\Delta_{g}}\omega\kappa\sin(\sqrt{\Delta_{g}}\omega t)\} .
\end{eqnarray}
Here, $\langle X(0)\rangle$ and $\langle P(0)\rangle$ are the expectation values of the position and momentum operators over the initial state of the bosonic field $|\psi(0)\rangle$,
$\mathrm{Cov}(X,P)=\langle\psi(0)|XP|\psi(0)\rangle-\langle\psi(0)|X|\psi(0)\rangle\langle\psi(0)|P|\psi(0)\rangle$. When the noise is absent, i.e., $\kappa=0$, it can be proven that the
local maxima of the inverted variance at the evolution time $\tau=2m\pi/(\sqrt{\Delta_{g}}\omega)$
is equal to $16g^{2}\omega^{2}\Delta_{g}^{-2}\tau^{2}\left\langle P(0)\right\rangle ^{2}/\left(\Delta X(0)\right)^{2}.$
It scales quadratically with $\tau$, and shows a divergent feature
when approaching the critical point, i.e., $F_g(\tau)\propto \Delta_{g}^{-3}\rightarrow\infty$.

\begin{figure}
\includegraphics[width=8.8cm]{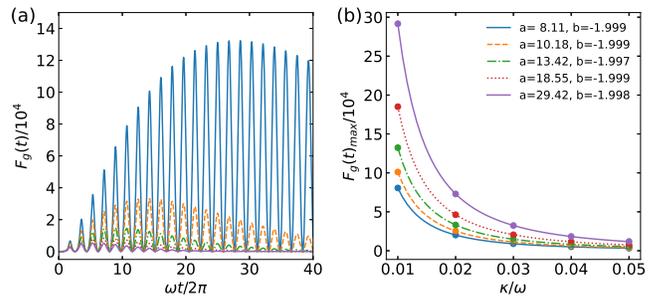}
\caption{The effect of the noise strength $\kappa$ on the precision of the criticality-enhanced metrology. (a) The inverted variance of the QRM in the vicinity of the critical point, e.g. $g=0.96$, when coupling with a thermal reservoir at zero temperature, for $100\kappa/\omega=1,~2,~3,~4,~5$. They correspond to the blue solid, orange dashed, green dash-dotted, red dotted, purple solid line, respectively. (b) The maximum of the inverted variance
as a function of the noise parameter $\kappa$. From the bottom to the top, the five curves correspond to $g=0.94,~0.95,~0.96,~0.97,~0.98,$ respectively. The dots are obtained from the master equation, while the curves are calculated by the numerical fitting. \label{fig1}}
\end{figure}

At zero temperature, the average photon number $\bar{n}$ equals zero.
In this case, we assume the same initial state $\left(|0\rangle+i|1\rangle\right)/\sqrt{2}$
as Ref.~\citep{Garbe2020PRL}. Figure~\ref{fig1} shows the inverted
variance of the QRM close to the critical point when coupling with
a thermal reservoir at zero temperature. As shown in Fig.~\ref{fig1}(a),
the dynamics of $F_{g}(t)$ manifest abundant phenomena. At first,
the envelope increases along with the increase of the time. Then, after
passing the maximum, it decreases and converges due to the noise.
It is in sharp contrast to the prediction when the noise is absent.
Furthermore, we also investigate the maximum values of $F_{g}(t)$
for different $\kappa$'s in Fig.~\ref{fig1}(b). As
shown by the numerical fitting, the dependence of the maximum inverted variance $F_g(t)|_\textrm{max}$ on $\kappa$ can be well described by $a\kappa^b$. For different $g$'s, although $a$'s are distinct, all $b$'s are non-exceptionally equal to $2$. 

\begin{figure}
\includegraphics[width=8.8cm]{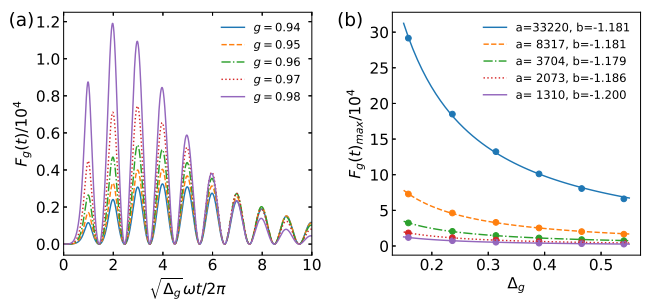}
\caption{The effect of the energy gap $\Delta_g$ on the precision of the criticality-enhanced metrology. (a) The inverted variance $F_{g}$ of the QRM when approaching the critical point, e.g. $g=0.94,0.95,0.96,0.97,0.98$, with the noise parameter $\kappa=0.05\omega$. It is plotted against the rescaled
time $\Delta_g\omega t/2\pi$ to highlight the behavior that
in the representation of the rescaled time $F_{g}$'s oscillate with
the same frequency for different $g$'s. (b) The maximum of $F_{g}$
vs $\Delta_g$ for $100\kappa/\omega=1,~2,~3,~4,~5$, which correspond to the curves from the top to the bottom. The dots are obtained from the master equation, while the curves are calculated by the numerical fitting.
\label{fig2}}
\end{figure}

In Fig.~\ref{fig2}(a), we plot the time dependence of $F_{g}(t)$
near the critical point for different $g$'s. Notice that the time is
rescaled by $\Delta_g\omega/2\pi$, which is approximately the oscillating
frequency of $F_{g}(t)$ for small $\kappa\ll\omega$. Similarly as in Fig.~\ref{fig1}(a), we
can see that $F_{g}(t)$ gradually increase from their initial values
$F_{g}(0)$ = 0 to the maximum values. 
Then, $F_{g}(t)$ begins to decrease and eventually vanishes in the
long-time limit. By theoretically obtaining $F_{g}(t)$ under different
$\kappa$'s, we plot the relation of the maximum of $F_{g}(t)$
with respect to $\Delta_g$ in Fig.~\ref{fig2}(b). We can see that $F_{g}(t)$
decays as $\Delta_g$ increases for a given $\kappa$. It indicates that
the noise hinders the metrological advantages in this framework. Again, we numerically fit the maximum of $F_{g}(t)$ as a function of $\Delta_g$, i.e., $a\Delta_g^b$. Despite of different $a$'s, all curves consistently decays with a nearly-identical $b\simeq-1.18>-2$, which is the noise-free case. This implies that in the noisy QRM, the inverted variance still diverges when approaching the critical point, i.e., $\Delta_g\rightarrow0$. In other words, by the scaling behavior, we theoretically prove that the criticality-based metrology is robust in the presence of single-photon relaxation.

\begin{figure}
\includegraphics[width=8.8cm]{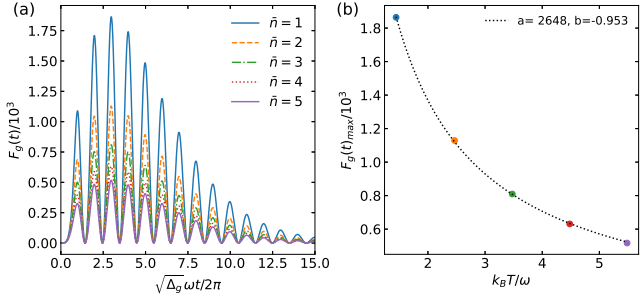}\caption{The effect of the temperature $T$ on the precision of the criticality-enhanced metrology. (a) The inverted variance of the QRM close to the critical point, at $g=0.96$, with $\bar{n}=1,~2,~3,~4,~5$ and $\kappa/\omega=0.05$. (b) The maximum inverted variance vs the temperature $T$. The balck dotted line is numerically fitted by $a(k_BT/\omega)^b$ with $a=2648$ and $b=-0.953$. \label{fig4}}
\end{figure}

We now investigate the influence of average photon number. As it can
be seen from Fig.~\ref{fig4}(a), the time-rescaled $F_{g}(t)$'s for
different $g$'s reach their local maxima at the same period while
maximum values of $F_{g}(t)$ decay as $\bar{n}$ increases. Again, we plot the maximum $F_{g}(t)$ vs the temperature $T$ in Fig.~\ref{fig4}(b). It's shown that the maximum $F_{g}(t)$ drops dramatically as $T$ rises. This observation reminds us that although the criticality-based metrology seems immune to the noise, the precision is very sensitive to the temperature. Hereafter, we will propose some method to relieve this disadvantage.

\begin{figure}
\includegraphics[width=8.9cm]{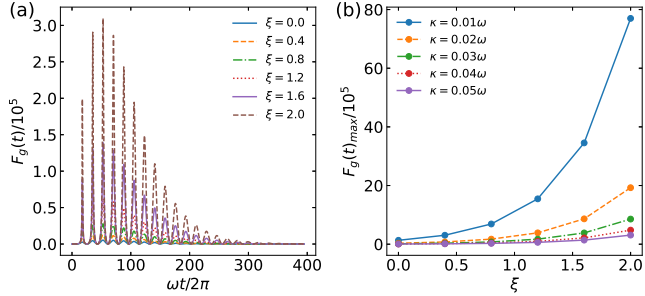}
\caption{Improving the precision by performing a squeezing operation on the initial state. (a)$F_{g}(t)$ versus time $t$, with different squeezing parameter $\xi$, when $g=0.96$ and $\kappa/\omega=0.05$. (b) The maximum of
$F_{g}$ at different $\xi$'s with $100\kappa/\omega=1,~2,~3,~4,~5$. \label{fig5}}
\end{figure}

According to Eq.~(\ref{eq:Fgt}), the measurement accuracy highly
depends on the initial state. This observation inspires us to perform
a squeezing operation, defined by a squeezing operator $S(\xi)=\exp[(\xi^{*}a^{2}-\xi a^{\dagger2})/2]$,
on the initial bosonic field state, where $\xi$ is the squeezing
parameter. As seen in Fig.~\ref{fig5}, the maximum of $F_{g}(t)$
is significantly increased as $\xi$ is enlarged. Although this gain can
not completely offset the influence of decoherence, it offers an alternative
way to improve the precision under decoherence.

\section{Two-Photon Relaxation}\label{sec:TwoPhoton}


In the above investigations, we explore the quantum metrology in the presence of single-photon relaxation. However, the quantum dynamics of the QRM with single-photon and two-photon relaxation are totally different \citep{Malekakhlagh2019PRL}. Inspired by this discovery, we study the quantum metrology in the QRM with two-photon relaxation, which can be described by the effective master equation $\dot{\rho}=-i[H_{\text{np}},\rho]+\gamma_{a}D[a^2]+\gamma_{h}D[a^{\dagger2}]$ \cite{Malekakhlagh2019PRL}.

\begin{figure}
\includegraphics[width=8.8cm]{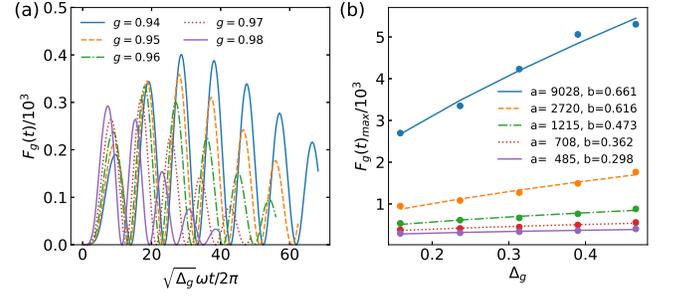}
\caption{The effects of the two-photon relaxation. (a) The inverted variance $F_{g}$ of the QRM when approaching the critical point, e.g. $g=0.94,0.95,0.96,0.97,0.98$, with the noise
parameter $\kappa=0.05\omega$. It is plotted against the rescaled
time $\sqrt{1-g^{2}}\omega t/\pi$. (b) The maximum of $F_{g}$
at different $g$'s with $100\kappa/\omega=1,~2,~3,~4,~5$. The dots are obtained from the master equation, while the curves are calculated by the numerical fitting.
\label{fig7}}
\end{figure}

In Fig.~\ref{fig7}(a), we investigate the dynamics of the inverted variance for different $g$'s under the influence of the two-photon relaxation. In contrast to the case with the single-photon relaxation, the quantum dynamics for different $g$'s do not oscillate with the same frequency for the rescaled time. In addition, as $g$ increases, the time to achieve the maximum becomes shorter and shorter. Moreover, we plot the maximum inverted variance as a function of $\Delta_g$ in Fig.~\ref{fig7}(b). Interestingly, the behavior is quite different from that for the single-photon relaxation in Fig.~\ref{fig1}(b). The maximum inverted variance increases monotonically as $\Delta_g$ is enlarged. We again numerically fit the data with the function $ax^b$ and thus obtain $a=9.03\times10^3$ and $b=0.661$ for $\kappa=0.01\omega$. As $\kappa$ increases, both $a$ and $b$ reduce dramatically. In other words, when we strengthen the noise, the inverted variance becomes smaller and shows weaker dependence on $\Delta_g$. This remarkable difference can be well described by the Schr\"{o}dinger equation with a non-Hermitian Hamiltonian \cite{Ai2021PRB,Dong2012}, when the relaxation rate is weak. When there is only single-photon relaxation, the non-Hermitian Hamiltonian is the original Hamiltonian $H_\textrm{np}$ plus $-i(\gamma_a a^\dagger a+\gamma_h aa^\dagger)/2$. Although the Hamiltonian is non-Hermitian, the energy spectrum is still equally-spaced, which is fundamentally required by the criticality-based metrology \cite{Chu2021PRL}. However, for the case with two-photon relaxation, the non-Hermitian Hamiltonian is the original Hamiltonian $H_\textrm{np}$ plus $-i(\gamma_a a^{\dagger2}a^2+\gamma_h a^2a^{\dagger2})/2$. Due to the non-linearity introduced by the two-photon relaxation, the energy spectrum is not equally-spaced and thus breaks down the underlying physical mechanism of the criticality-based metrology.

\section{Conclusion}\label{sec:Conclusion}

We have investigated the impact of decoherence on the criticality-based
quantum metrology in the QRM, showing that the achieved precision
still diverges when approaching the criticality.
In particular, we consider the
single-photon relaxation described by the Lindblad-form master equation. By the semiclassical equation of motion, we obtain the analytical solution for the precision characterized by the maximum inverted variance. We have shown that the precision is very sensitive to the temperature. As a remedy, we propose to improve the precision by performing the squeezing operation on the initial state. Furthermore, we also investigate the performance of the metrology in the presence of two-photon relaxation. In contrast to the single-photon relaxation, the inverted variance do not oscillate against the rescaled time at the same frequency for different $g$'s. More strikingly, the maximum inverted variance shows a completely-different tendency when closing to the quantum phase transition.

This work is supported by Beijing Natural Science Foundation under Grant No.~1202017 and the National Natural Science Foundation of China under Grant Nos.~11674033, 11505007, and Beijing Normal University under Grant No.~2022129.
%
\end{document}